# A Stochastic Singular Vector Based MIMO Channel Model for MAC Layer Tracking

Tim W. C. Brown and Patrick C. F. Eggers

*Abstract*— A novel stochastic technique is presented to directly model singular vectors and singular values of a multiple input multiple output (MIMO) channel. Thus the components modeled directly in the eigen domain can be adapted to exhibit realistic physical domain behavior when assembled. The model exploits natural paths of eigenmodes, such that a simple Doppler filter generator process can be used. Furthermore, it is possible to directly manipulate the singular vector dynamics in a way that an unrealistic "stress channel" can be modeled in the eigen domain. This is particularly useful for testing the eigenmode channel tracking ability internal to a communication device such as a modem, where impairments in tracking will cause interference between eigenmodes. The model can also facilitate mode tracking testing, as it directly produces tracked (untangled) eigenmodes, providing the narrowest possible singular vector Doppler spectra and consequently lowest required update rates of each eigenmode. This singular vector based model targets testing of the eigen domain functionality of MIMO modems/devices (i.e. an apparatus focus) without the need for including the decomposition stages.

## I. Introduction

MULTIPLE input multiple output (MIMO) communications have become an attractive option for cases where spectrum efficiency is desirable [1]. Several channel models in the physical domain have been developed in recent times in order to ascertain the level of scattering richness [2] and signal to noise ratio [3]-[10], both of which are highly important to properly predict the Shannon capacity. These models have addressed the need to extend beyond the traditional idea of independently identically distributed (i.i.d) Rayleigh channels. For example they accommodate features such as the keyhole effect, antenna interactions, joint angle of arrival (AOA) and angle of departure (AOD) distributions at the transmitter and receiver. These multipath environment factors and others will determine the inter-dependence of the branches in the MIMO channel, which will impact the channel capacity. Previous stochastic based MIMO models have included the correlation based Kronecker model devised in [11] and experimentally evaluated in [12]. However, subsequent deficiencies were identified, which led to extending the model towards semi eigen based models [13]-[15]. All of these channel models have specifically focused on modeling realistic eigenvalue behaviors (i.e. channel gains), which directly impact system capacity. None of these models, however, include direct eigenmode modeling of the time variant singular vector dynamics (i.e. the orthogonal channel matching) as detailed in section II of this paper. These are required to evaluate modem performance of practical MIMO terminals. The eigenmodes are defined as the orthogonal modes by which data can be transmitted using the corresponding singular vectors, where an eigen state is defined as an instantaneous state of an eigenmode.

In order for the capacity potential of a channel to be reached, it is important to have a device employing MIMO such as a modem with a transceiver that can reliably track the changes in channel states and corresponding eigenmodes [16][17]. The non stationarity of MIMO channels has also been studied with measured and modeled results available in the literature. One metric that has commonly been used is the wide sense stationarity (WSS), where it is also analyzed between MIMO branches with non-consistent characteristics [18]. Another metric applied to the channel stationarity is the correlation matrix, analyzed through measurement and models [19]-[21], such that a threshold correlation can be identified due to change in channel state at a given displacement of the receiver. Finally the F-eigen ratio earlier defined in [22] has been proposed as a measure of degradation due to change in channel state. All these metrics focus on the channel state at the physical layer (PHY).

At the media access control (MAC) layer, singular vectors are time varying with significantly different rates and Doppler spread compared to PHY, where their change in eigen state is not directly linked to the change in physical channel state [23]. The dynamics of singular vectors also have more than one possible solution in order to assemble the same physical channel. A model is therefore needed for product testing this singular vector tracking at MAC layer, where the singular vector dynamics can be directly controlled, which provides the opportunity to add "stress" into the dynamics and therefore directly test the extreme limits of the singular vector tracking. Singular vectors could be modeled by using available physical layer models but it would then require several iterations to decompose to the corresponding singular vectors and it would not be possible to directly add "stress" to assist in product testing.

In addressing the need to directly model singular vector dynamics, this paper presents a novel stochastic channel model, which extends beyond previous eigen based models and the controlling parameters are entirely applied in the eigen domain. The challenges of modeling first and second order statistics are met when it is necessary to ensure the singular vectors are unitary and orthogonal. A further challenge is that a realistic physical channel could then be assembled with realistic channel Doppler spectra such as the Classical case in [7]. This paper explains the modeling procedure and how the singular vectors can be manipulated to introduce "stress" into the channel.

Section II considers how changes in singular vectors and eigen state can impact channel capacity through defining the signal to interference ratio between eigenmodes while also it defines a set of model classes used in this paper. Sections III and IV will detail how both first and second order statistics of both singular vectors and singular values can be modeled stochastically. This is validated by assembly into realistic physical channels. Finally Sections V and VI will conclude by demonstrating applications of the model.

## II. SINGULAR VECTOR STATES AND MODEL CLASSES

The singular vectors and singular values are here defined by taking the singular value decomposition (SVD) of a physical MIMO channel $\mathbf{H}$, which can be represented in the 2x2 case as follows [24]:

$$\mathbf{S} = \mathbf{U}^H \mathbf{H} \mathbf{V} = \begin{bmatrix} \mathbf{u}_1^H \\ \mathbf{u}_2^H \end{bmatrix} \begin{pmatrix} h_{11} & h_{21} \\ h_{12} & h_{22} \end{pmatrix} [\mathbf{v}_1 \ \mathbf{v}_2] = \begin{pmatrix} s_1 & 0 \\ 0 & s_2 \end{pmatrix} \quad (1)$$

where each of the singular values, $s_i$ have corresponding unitary and orthogonal singular vectors, $\mathbf{u}_i$ and $\mathbf{v}_i$, at each end. The corresponding eigenvalues, $\lambda_i = |s_i|^2$. It should be noted that the singular vectors $\mathbf{V}$ are equal to eigenvectors, which could be generated from eigenvalue decomposition.

Table 1 compares the proposed singular vector model introduced in this paper with existing stochastic MIMO channel models. The correlation based models consider the fixed correlation matrix at the transmitter, $\mathbf{R}_{Tx}$ and receiver, $\mathbf{R}_{Rx}$ independently, while the i.i.d channel component, $\mathbf{H}_{iid}$, is time variant. These models have several deficiencies, while also there is no direct inclusion of singular vector behavior. Virtual channel based modeling pre-defines the fixed beamforming vectors at both ends, $\mathbf{A}_{Tx}$ and $\mathbf{A}_{Rx}$, assuming a uniform linear array, though they are not comparable to realistic singular vectors in a channel because of this constraint. Unlike the correlation based models, the time variant channel component consists of discrete Fourier transform elements, $\mathbf{H}_D(t)$. Eigen based models were initially derived by Weichselberger [13] and further developed by others as referenced in Table 1. However, the singular vectors in this model are created by a direct decomposition of the fixed correlation matrices at both ends (thus taking several iterative steps) in order to form fixed vectors at each end, $\mathbf{U}_{Tx}$ and $\mathbf{U}_{Rx}$. In this model only the singular values (channel gains) are indirectly controlled by $\mathbf{H}_{iid}$ and the fixed control parameters in matrix $\mathbf{\Omega}$ (based on the structure of scatterers in the physical domain).

| Model Base | Input(*t*), method, domain | Composed or decomposed domain |
|---|---|---|
| Correlation [11][12] | $\mathbf{H}_{iid}(t)$ <br> $\mathbf{H}_{kron} = \mathbf{R}_{Tx}\mathbf{H}_{iid}(t)(\mathbf{R}_{Rx})^H$ | $\mathbf{U},\mathbf{S},\mathbf{V} = \mathrm{SVD}(\mathbf{H}_{kron})$ |
| Virtual channel [14] | $\mathbf{H}_D(t)$ <br> $\mathbf{H}_{virtual} = \mathbf{A}_{Tx}\mathbf{H}_D(t)\mathbf{A}_{Rx}^H$ | $\mathbf{U},\mathbf{S},\mathbf{V} = \mathrm{SVD}(\mathbf{H}_{virtual})$ |
| Eigen base [13] ([25]-[28],[15]) | $\mathbf{H}_{iid}(t)$ <br> $\mathbf{H}_{eigenbase} = \mathbf{U}_{Tx}(\mathbf{\Omega} \circ \mathbf{H}_{iid}(t))\mathbf{U}_{Rx}^H$ | $\mathbf{U},\mathbf{S},\mathbf{V}$ <br> $= \mathrm{SVD}(\mathbf{H}_{eigenbase})$ |
| Singular vector (proposed) | $[\mathbf{U}(t_0),\mathbf{V}(t_0)]_\perp \quad a_{VD_{u,v,s}}(t)$ <br> $\mathbf{u}_1(t) = a_{VD_u}(t)$ <br> $\mathbf{v}_1(t) = a_{VD_v}(t)$ <br> $\mathbf{U}(t_{n+1}) = \mathbf{T}_u(\mathbf{u}_1(t_n))\mathbf{U}(t_n)$ <br> $\mathbf{V}(t_{n+1}) = \mathbf{T}_v(\mathbf{v}_1(t_n))\mathbf{V}(t_n)$ <br> $\mathbf{S}(t) = \mathbf{I}a_{VDs}(t)$ | $\mathbf{H}_{SingularVector}$ <br> $= \mathbf{U}(t)\mathbf{S}(t)\mathbf{V}^H(t)$ |

**Table 1 - Summary of existent and newly proposed correlation, virtual channel and eigen based models. Notation (*t*) indicates where time dependence is introduced and *n* the current state (time) step**

The novelty of the model proposed in section III lies in the time variant and orthogonal singular vector matrices denoted as $[\mathbf{U}(t),\ \mathbf{V}(t)]_\perp$. The vectors and singular values $\mathbf{S}(t)$ include virtual Doppler (VD) noise generators, represented here in symbolic time domain $a_{VDu,v,s} = \mathcal{F}^{-1}[A_{fdu,v,s}]$. The frequency domain filters $A_{fdu,vs}$, are used in equation (4) for singular vectors, while for singular values in equation (9). $\mathcal{F}^{-1}[]$ is an inverse Fourier transform function. The remaining vectors in the matrices, $\mathbf{U}(t_n)$ and $\mathbf{V}(t_n)$, are generated iteratively from state $n$ to $n+1$ using Householder transforms, $\mathbf{T}_u$ and $\mathbf{T}_v$ defined in equation (A.13) in Appendix A. This assumes the first instantaneous states, $\mathbf{U}(t_0)$ and $\mathbf{V}(t_0)$, are known. If required, the singular vectors and values can be assembled into a full physical channel, as shown in the third column of Table 1. If the time variant channels within $\mathbf{H}_{SingularVector}$ are analysed in the Doppler domain they will show the desired spectra, thus validating that the singular vector and value dynamics have been correctly modeled. The proposed model therefore does not require several iterations using an SVD algorithm to model the singular vectors unlike prior models, thus proving its useful application directly at MAC layer.

In equation (1) there are two eigenmodes, $\mathbf{u}_1^H\mathbf{H}\mathbf{v}_1 = s_1$ and $\mathbf{u}_2^H\mathbf{H}\mathbf{v}_2 = s_2$. The physical channel, $\mathbf{H}$, is applied in the physical domain as $\mathbf{y} = \mathbf{H}\mathbf{x} + \mathbf{n}$, where $\mathbf{x}$ and $\mathbf{y}$ are the input and output data streams respectively and $\mathbf{n}$ is additive noise. If the system uses the singular vectors corresponding to outdated channel state information (CSI) when the channel changes, then the eigenmode transmissions become non-orthogonal. Therefore the relevant measure for each received eigenmode becomes the signal to interference and noise ratio (SINR). This paper will focus on the impact of inter-mode interference itself, via the signal to interference ratio (SIR), defined in equation (3). To represent a case where the singular vectors are no longer tracking orthogonal channels, notations $\hat{\mathbf{U}}$ and $\hat{\mathbf{V}}$ can be applied to equation (1) to derive the new non diagonal $\mathbf{S}$ matrix, $\hat{\mathbf{S}}$ as follows[1]:

$$\hat{\mathbf{S}} = \hat{\mathbf{U}}^H \mathbf{H} \hat{\mathbf{V}} = \begin{pmatrix} \hat{s}_{11} & \hat{s}_{21} \\ \hat{s}_{12} & \hat{s}_{22} \end{pmatrix} \quad (2)$$

With this scenario, the eigenmodes with interference are represented as $\hat{\mathbf{u}}_1^H \mathbf{H} \hat{\mathbf{v}}_1 \equiv [\hat{s}_{11}\ \hat{s}_{21}]$ and $\hat{\mathbf{u}}_2^H \mathbf{H} \hat{\mathbf{v}}_2 \equiv [\hat{s}_{22}\ \hat{s}_{12}]$. Correspondingly $SIR_1 = |\hat{s}_{11}|^2/|\hat{s}_{21}|^2$ and $SIR_2 = |\hat{s}_{22}|^2/|\hat{s}_{12}|^2$. For an $N$x$N$, $N$x$M$ or $M$x$N$ channel where $N < M$, $SIR_i$ is defined in general where $i \in [1\ N]$ as:

$$SIR_i = \frac{|\hat{s}_{ii}|^2}{\left|\sum_{j=1, j \neq i}^{N} \hat{s}_{ji}\right|^2} \quad (3)$$

If there is perfect CSI then $\hat{s}_{ii} = s_i$ and $\hat{s}_{ji} = 0$, thus $SIR_1 = SIR_2 = \infty$ in a 2x2 channel. The SIR impact on outdated singular vectors is greatest by the most commonly used SVD algorithm itself, LAPACK [28], which will force a sorting of the singular values into descending order according to their gains. This is exemplified in Figure 1 for a 2x2 (i.i.d) Rayleigh MIMO channel, with a classical Doppler spectrum and maximum Doppler shift $f_d$.

At the time instance of $0.4/20f_d$, $s_1$ and $s_2$ come close together. If the singular values followed their natural path, then $s_2$ would become larger than $s_1$ and so the algorithm will forcibly re-order the two singular values and their corresponding singular vectors to maintain the descending order. Therefore $\mathbf{v}_1$ is swapped with $\mathbf{v}_2$, while also $\mathbf{u}_1$ is swapped with $\mathbf{u}_2$ and all variables are assigned their new values as a result of the swap to maintain orthogonality. This is shown where $|\mathbf{u}_1(t).\mathbf{u}_1(t+\delta t)|$ in Figure 1 (b) instantaneously falls to a low value after a swap takes place.

It is necessary to use a small enough sample spacing, $\delta t = 1/f_s$, to see the swap happen, thus sampling frequency $f_s = 20f_d$. In the SIRs in Figure 1 (c), it is assumed that at $t = 0$ the channel has been learned with perfect CSI (resulting in infinite SIR). However, from this point on the weights for singular vector matrix $\mathbf{U}$ are fixed, while $\mathbf{V}$ are updated, therefore $\hat{\mathbf{u}}_1^H(0)\mathbf{H}\hat{\mathbf{v}}_1(t) \equiv [\hat{s}_{11}(t)\ \hat{s}_{21}(t)]$ and $\hat{\mathbf{u}}_2^H(0)\mathbf{H}\hat{\mathbf{v}}_2(t) \equiv [\hat{s}_{22}(t)\ \hat{s}_{12}(t)]$.

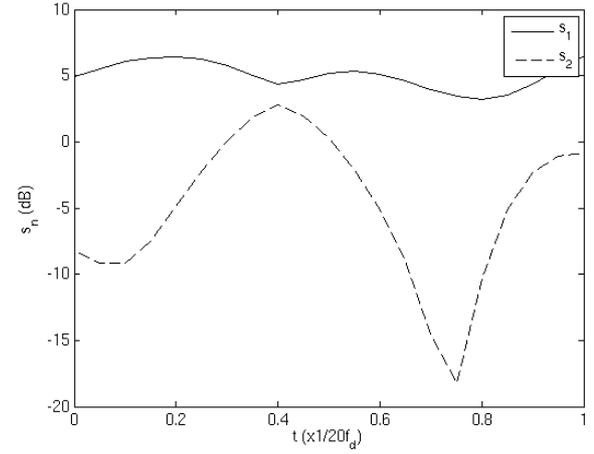

(a)

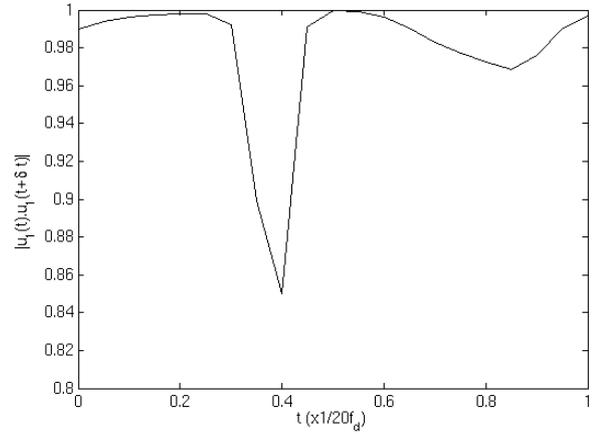

(b)

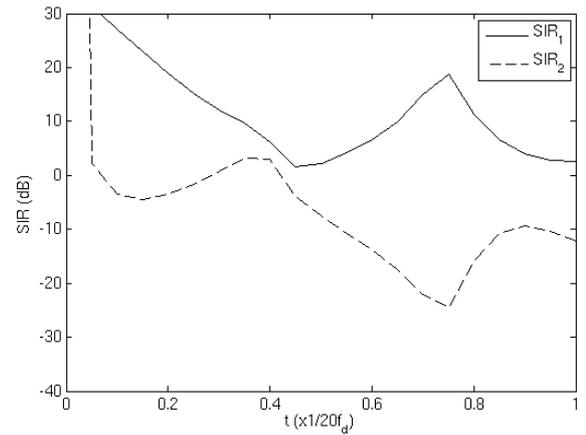

(c)

**Figure 1 - Comparison of (a) instantaneous singular values, (b) the state transition of the singular vector $\mathbf{u}_1$ and (c) the resultant SIRs of a 2x2 MIMO channel.**

Singular vector swaps can be removed by using mode untangling algorithms such as in Browne [23]. Therefore SIRs would not drop so rapidly. The difference in this case is that it is variable as to which eigenmode has the largest diversity order. It is also useful to note that if water filling [24] were applied, then this would also impact the corresponding SIRs between eigenmodes. Therefore a MIMO device will have to maintain a frequent update rate in order that it can continue to maintain high SIRs. The purpose of the channel model presented in this paper is to deliberately create "stress" instances that will cause such SIRs to fluctuate in different ways, which would allow assessment of the modem's tracking performance. The implementation of deterministic and semi-deterministic class I to class IV models are summarized in Appendix A due to their simplicity.

---

[1] Non orthogonal decomposition components and values are denoted here by a *hat* i.e. $\hat{s}$ to distinguish it from a true orthogonal singular value $s$.

Finally there is a class V model (which is the same as the model proposed in Table 1). This is entirely stochastic in the eigen domain with representative singular vector and singular value dynamics for a real physical channel with details in Section III. The different model classes in this paper can be grouped into three families as follows:
1. **Family I** Deterministic and semi-deterministic (Classes I, II and III): $s_i$ constant for each eigenmode.
2. **Family II** Semi-deterministic (Class IV): $\sum \lambda_i$=constant thus a fixed total power gain while applying realistic vector dynamics.
3. **Family III** Stochastic (Class V): Fully stochastic model (singular vectors and values) but setting singular value distributions for certain test cases.

### III. SPACE VARIANT MODEL PROCEDURE

This section will describe the steps in the modeling procedure in general for an $N \times M$ channel. To generate the singular vectors and singular values with correct Doppler spectra, auxiliary parameters have been derived, dependent on $N$ and $M$. Different spectra can be generated, but for the scope of this paper, Classical Rayleigh/Rice spectra [7] will be used (corresponding to omni directional scattering).

#### A. Step 1 - Generate singular vectors $\mathbf{u}_1$ and $\mathbf{v}_1$ with Doppler filtering

The greatest challenge the model is the virtual Doppler noise generators to make realistic first and second order statistics of the singular vector elements, while at the same time maintaining their unitary and orthogonal properties. In the PHY domain channels are modeled by generating Monte Carlo samples, which are either filtered in the time domain (by autocorrelation) or in the frequency domain (digital filtering). In both cases, the samples' magnitude is attenuated, which destroys the unitary properties of singular vector elements. A frequency domain filtering approach is used where the Doppler components, or tones, have set magnitudes defined by a filter function, which when summed together work as a virtual Doppler noise generator. The filter function, $A_{\text{fduv}}$ has been tailored to generate virtual Doppler noise where the vector elements follow their natural path (see Appendix B):

$$A_{\text{fduv}}(f) = \begin{cases} \dfrac{1}{0.6 N_{\text{sam}} \sqrt{1+K_f \left|\dfrac{f}{f_d}\right|}} & -\dfrac{f_d}{4} < f < \dfrac{f_d}{4} \\ A_{\text{uv}} & f = 0 \\ 0 & \text{Otherwise} \end{cases} \quad (4)$$

where $N_{\text{sam}}$ is the number of samples in the time domain. $A_{\text{uv}}$ and a factor of 0.6 are auxiliary parameters. For an $N \times N$ singular vector matrix, $A_{\text{uv}} = 1/\sqrt{N}$, derived by simulation as an upper limit for all Doppler components close to the asymptote at zero frequency, thus $A_{\text{fduv}}(f) \leq A_{\text{uv}}$. Any Doppler components above $A_{\text{uv}}$ would prevent the singular vectors from holding unitary nature. The Rice factor, $K_f$ is the ratio of the line of sight to scattered received power, $K_f = P_{\text{LOS}}/P_{\text{Scatter}}$ [29][2], which is zero for the Rayleigh case. For realism of an assembled equivalent physical channel $\mathbf{H}$, its Doppler spectrum should be bounded between the maximum Doppler shift, $\pm f_d$. The corresponding Doppler spectrum appears due to a frequency domain convolution of the spectra of the individual $\mathbf{u}_i$, $s_i$ and $\mathbf{v}_i$ elements. For simplicity, a $\pm f_d$ constraint is imposed on each $s_i$ element while the $\mathbf{u}_i$ and $\mathbf{v}_i$ elements have a constraint within $\pm f_d/4$. The 0Hz component for the singular vectors is substantially larger than the non zero Doppler components, dictated by the factor $0.6N_{\text{sam}}$, adding only little to the total spectral broadening when convolution happens. Thus, this combination of Doppler spectra in the eigen domain results in a convolution yielding a physical channel with a plausible $\pm f_d$ constrained Doppler spectra. The factor $0.6N_{\text{sam}}$ must also be small enough to enable sufficiently variable singular vector dynamics, which were derived by trial and error in simulation.

Note that with the Doppler spectra on the singular vectors, the fading is slow and as desired comparable to that generated after SVD untangling [23], this is illustrated in Figure 8 in Appendix B. Equation (4) is applied to generate the virtual Doppler noise by summing the Doppler components. Each element of the first column of any sized singular vector matrix $\mathbf{U}$, denoted as $u_{1i}$:

$$u_{1i}(t_n) = \sum_{p=1}^{N_{\text{freq}}} A_{\text{fduv}}(f_p) e^{j\left(\dfrac{2\pi n}{S_f}\dfrac{f_p}{f_d}+\phi_{p,i}\right)} \quad (5)$$

where $i \in [1 \; N-1]$ and $N$ is the number of elements in the column. The sampling frequency $f_s > 2f_d$ thus $S_f > 2$ to meet Nyquist criteria, while the Doppler components $f_p \in f_d \cdot [-0.255 \; 0.255]$. This restricted range was also derived by trial and error in simulation to yield the desired Doppler spectra in the physical domain. The number of Doppler components derived must also meet the criteria, $N_{\text{freq}} = N_{\text{sam}}/30$. A smaller number of Doppler components will cause the magnitude of each sample to be too small or insignificant, while a larger number will cause the singular vectors to be no longer unitary. Each Doppler component, $f_p$ has a uniformly distributed random phase, $\phi_{p,i}$. To avoid startup effects it is necessary to discard the first 20% of samples. The magnitude $|u_{1N}|$, must be determined as follows to ensure a unitary singular vector:

$$|u_{1N}| = \sqrt{1-|u_{11}|^2-|u_{12}|^2-|u_{13}|^2....-|u_{1(N-1)}|^2} \quad (6a)$$

---

[2] The model could be equally adapted to Nakagami distributions [29], via suitable Doppler filters

$$\sum_{i=1}^{N-1}|u_{1i}|^2 \leq 1 \quad (6b)$$

The derived virtual Doppler noise generators ensure the sum constraint in equation (6b) is only violated for a small fraction of the time (< 1%). Where this violation does occur, a cap can be applied to the preceding elements at that time instant while $u_{1N}$ is set to zero. The capping causes little disruption to the Doppler spectra. For high $N$ this can be more vulnerable to happening where care would be necessary to ensure unitary conditions are maintained.

The phase for $u_{1N}$ must follow a suitable trajectory, which will ensure the Doppler Spectrum is comparable to any other $i^{th}$ element $u_{1i}$. Due to the slow variation in the fading it is possible to generate phase trajectory independent of $|u_{1N}|$ and obtain the desired Doppler spectra. This would be not the case if a wider Doppler spectrum was used for the singular vectors. The independent trajectory is generated with a new set of random phases, $\phi_r$:

$$e^{j\theta_N}(t_n) = \frac{\sum_{r=1}^{N_{freq}} A_{fduv}(f_r) e^{j\left(\frac{2\pi n}{S_f}\frac{f_r}{f_d}+\phi_r\right)}}{\left|\sum_{r=1}^{N_{freq}} A_{fduv}(f_r) e^{j\left(\frac{2\pi n}{S_f}\frac{f_r}{f_d}+\phi_r\right)}\right|} \quad (7)$$

After the startup effects are removed by discarding the first 20% of samples $u_{1N}$ is thus derived:

$$u_{1N}(t_n) = |u_{1N}(t_n)| e^{j\theta_N}(t_n) \quad (8)$$

The same process can be used to create singular vector elements $v_{11}$ through to $v_{1N}$.

### B. Step 2 – Generate remaining singular vectors

Once the singular vectors $\mathbf{u}_1$ and $\mathbf{v}_1$ have been created, in the case of 2x2, the remaining singular vector elements can be easily derived as $|u_{11}| = |u_{21}|$, $|u_{22}| = |u_{12}|$ and $u_{11}u_{21}^* = u_{12}u_{22}^*$. For higher order channels, this is not possible and it is necessary to apply the Householder transform in equation (A.12) in Appendix A, which can be used for any size square singular vector matrix. Therefore the iterative process used in general for $N$x$N$ singular vector matrices is as follows:

1. Generate only the first initial state eigenvectors using an SVD algorithm from a randomly generated $N$x$N$ channel sample, $\mathbf{H}(t_0) = \mathbf{U}(t_0)\mathbf{S}(t_0)\mathbf{V}(t_0)$.
2. Calculate Householder transition matrix, $\mathbf{T}_u(t_1)$, using equation (A.13) and eigenvector states $\mathbf{u}_1(t_1)$ and $\mathbf{u}_1(t_0)$.
3. Calculate singular vector matrix, $\mathbf{U}(t_1) = \mathbf{T}_u(t_1)\mathbf{U}(t_0)$.
4. Repeat stages 2 and 3 iteratively for every state transition $n$, such that $\mathbf{U}(t_n) = \mathbf{T}_u(t_n)\mathbf{U}(t_{n-1})$.
5. Repeat stages 2 to 4 to generate vectors $\mathbf{V}(t_n)$.

### C. Step 3 – Generate singular values

The singular values, do not need to be unitary and in this paper they are based on a Classical filter function within the physical Doppler bounds, $\pm f_d$, as follows [29]:

$$A_{fds}(f) = \begin{cases} \dfrac{1}{\sqrt{\left|1-\left(\dfrac{f}{f_d}\right)^2\right|}} & -f_d < f < f_d \\ 0 & \text{Otherwise} \end{cases} \quad (9)$$

Note however, that applying a Classical Doppler spectra to the singular values is not representative of any real channel. However, it is not the intention in this model to do so, it is the intention to create "stress" on the MAC layer so different Doppler shifts can be set to vary this stress, while still assembling a plausible physical channel. Hence realistic channel scenarios are being tested, but still there is the ability to directly control the model at MAC layer. Each singular value, $s_i$, can be computed, where frequency or time domain filtering can be chosen as at PHY layer, though frequency domain filtering is chosen here:

$$s_i(t_n) = \frac{\sqrt{K_f} e^{j\frac{2\pi n}{S_f}}}{\sum_{q=1}^{N_{freq}} A_{fds}(f_q)} + \sum_{q=1}^{N_{freq}} A_{fds}(f_q) e^{j\left(\frac{2\pi n}{S_f}\frac{f_q}{f_d}+\psi_{q,i}\right)} \quad (10)$$

which includes the Ricean component and another set of uniformly distributed random phases $\psi_{q,i}$ different for each singular value. Again it is necessary to discard the first 20% of samples due to startup effects. It is also recommended to set the sampling frequency range in this case from $-4f_d$ to $4f_d$ well above the Nyquist criteria, therefore $S_f = 8$ or $f_s = 8f_d$. If desired, $N$-1 ratios can be set between the mean of the singular values thus offsetting the distributions of the singular values:

$$s_{\text{ratio }j} = \frac{|s_{j+1}|}{|s_j|} \quad (11)$$

where $j \in [1\ N-1]$. Whether or not the ratios are set, the singular values must be normalized to meet the following mean value criteria for $N$ eigenvalues [30]:

$$\overline{\lambda_1} + \overline{\lambda_2} + \ldots \overline{\lambda_N} = \overline{|s_1|^2} + \overline{|s_2|^2} + \ldots \overline{|s_N|^2} = N^2 \quad (12)$$

### D. Step 4 – Assemble the physical channel

If there is a need, the equivalent physical channel $\mathbf{H}(t_n) = \mathbf{U}(t_n)\mathbf{S}(t_n)\mathbf{V}^H(t_n)$ can be assembled for each sample $n$.

### E. Extra notes on higher order MIMO channels

For an $N$x$M$ model, the following further points need to be noted:

- In step 2, the Householder transform will work for $N$x$N$ or $M$x$M$ square matrices.
- For $\mathbf{H} \in \mathbb{M}_{N \times M}$ where $N > M$, vectors $\mathbf{V} \in \mathbb{M}_{M \times M}$ and $\mathbf{U} \in \mathbb{M}_{N \times N}$, can be generated by Householder transforms. $\mathbf{S} \in \mathbb{M}_{N \times M}$ where rows $\mathbf{S}_{M+1..N} = 0$. The $M$ singular values $s_1..s_M$ can then be inserted diagonally into rows (or columns) $1..M$.
- If instead $M > N$, again $\mathbf{V} \in \mathbb{M}_{M \times M}$ and $\mathbf{U} \in \mathbb{M}_{N \times N}$, but $\mathbf{S} \in \mathbb{M}_{N \times M}$ where columns $\mathbf{S}_{N+1..M} = 0$. The $N$ singular values $s_1..s_N$ can then be inserted diagonally into rows (or columns) $1..N$.
- The singular values must be normalized as in equation (12) such that $\sum \lambda_i = MN$ [30].

## IV. MODEL VALIDATION

To validate the implemented model, first a 2x2 assembled i.i.d Rayleigh channel is used. Figure 2 shows the gradient of 10dB per decade as is expected for a Rayleigh distribution. Figure 3 shows the expected Classical Doppler spectrum.

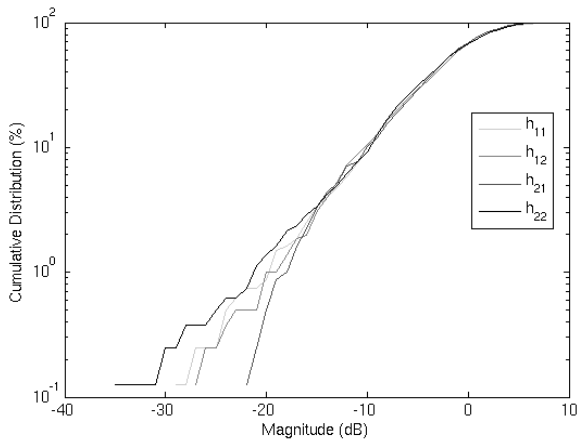

**Figure 2 - Validation of the cumulative distribution of the 2x2 model**

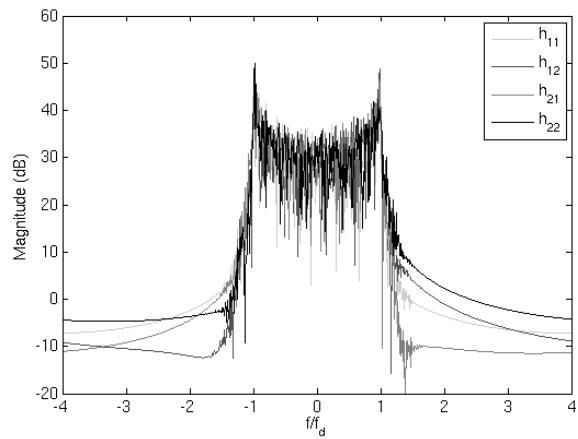

**Figure 3 - Validation of the Doppler Spectrum of the 2x2 model**

The Doppler skirts (or noise) at frequencies outside $\pm f_d$ region are more than 40dB down from the maximum. This provides a plausible output comparable with the Doppler Spectrum of real measured channels in omnidirectional scattering scenarios. In the case of the 4x4 channel, the 16 curves in the cumulative distribution plots in Figure 4 are again in good agreement with the expected 10dB/decade Rayleigh distribution. In Figure 5, only the Doppler spectra of assembled channels, $h_{11}$, $h_{22}$, $h_{33}$ and $h_{44}$ are plotted for clarity. The other remaining 12 channels have comparable Doppler spectra.

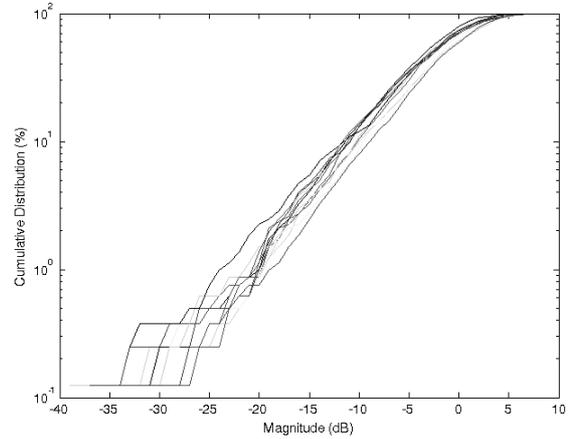

**Figure 4 - Validation of the cumulative distribution of the 4x4 model**

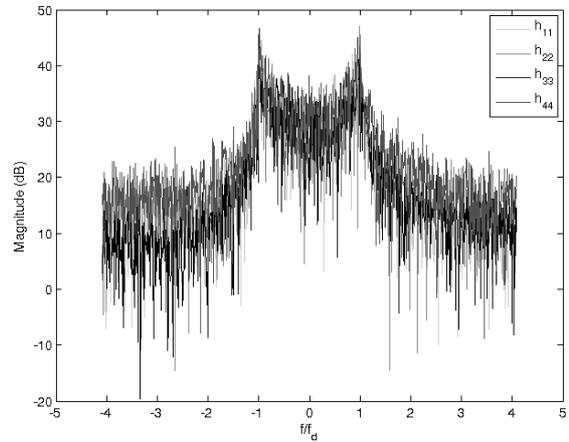

**Figure 5 - Validation of the Doppler Spectrum of the 4x4 model**

The out of Doppler band noise in the 4x4 case, though over 30dB down, is higher than that of the 2x2 case. This is due to a limitation of the singular vector elements generated by the Householder transform for a larger matrix, which causes the elements of the $\mathbf{U}$ and $\mathbf{V}$ matrices have higher noise at higher Doppler frequencies and consequently the $\mathbf{H}$ matrix. This does therefore indicate a limitation the model will have for higher order $N$x$N$ channels, though maintaining the strength of being able to manipulate the channel directly in the eigen domain. Compiled MATLAB code implementation of the class V model is available at the following web link: www.tim.brown76.name/SingularVectorMIMO.

## V. MODEL APPLICATION FEATURES AND EXAMPLES

The main feature of the class V model in the numeric complexity sense is detailed in Appendix B. The Appendix clarifies the ability of the model to directly generate and control the eigen domain matrices **U**, **S** and **V** (which would otherwise require several iterations from applying SVD in the physical domain). As a useful example of how the class V model can be manipulated to increase "stress" on the eigen tracking by directly controlling the singular vector dynamics, it is decided here that the singular vectors in **U** will be forcibly swapped every two samples by applying equation (A.3), while those in **V** will not be changed. When the channel **H** is assembled in such a case, it will be no longer realistic, with Doppler components greater than $f_d$. However, the point of the channel is to cause periodic occurrences of singular vector swaps so as to apply stress to the eigenmode tracking. This can only be achieved by modeling first in the eigen domain. In the development stages of a tracking device, the eigen domain components of this model can be used directly. If a full modem or communication device requires testing, the same channel conditions can be imposed via assembling the model to the physical domain and using a channel emulator. In this example, the simulated 4x4 channel model is used where only $SIR_1$ will be analyzed to show the model's function.

In Figure 6, $SIR_1$ is compared in three different scenarios and the SIR is plotted in the same manner as in Figure 1 (c). It is shown that for the first curve, the SIR is plotted as a result of tracking the singular vectors generated by the model, where it does not change rapidly because the natural path of the eigenmodes are followed. However, if the swaps are forced, the SIR changes very rapidly every two samples. This is comparably worse than when the channel is composed, then decomposed by LAPACK (where the eigenmodes will be arranged into descending order and fewer swaps occur). This is one example of how the channel model can be forcibly changed into unrealistic states for purposes of testing eigen tracking.

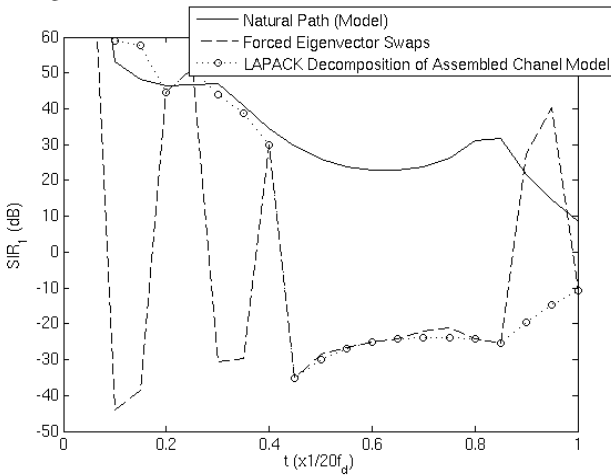

**Figure 6 - Comparison of the SIR Class V model using the natural path and with forced singular vector swaps increasing the "stress" on the channel**

## VI. CONCLUSION

A stochastic direct singular vector based MIMO channel model has been presented. The direct modelling of singular vectors and singular values is validated by assembling a plausible model in the physical domain. The benefit of the presented a model, is avoiding the need for several iterations using SVD algorithms from a physical MIMO channel model. This facilitates fully controlled eigenmode behavior, particularly with respect to the singular vectors. Results show that it is a suitable model for comparative development of eigenmode operation algorithms/protocols and possible conformance testing of singular vector tracking of MIMO communication devices. While this paper presents channel operations within the eigen domain, there is no principal hindrance to assemble the created eigen channel to a physical representation, which can be used in channel emulators if required. The model also has a secondary benefit of assembling multiplex rich channel models, which may be difficult to construct directly in the physical domain.

### APPENDIX A – IMPLEMENTATION OF CLASS I TO IV MODELS

The class I to IV models are summarized in the following sections, where also example MATLAB code implementations of the models for both 2x2 and 4x4 channels are available at www.tim.brown76.name/SingularVectorMIMO/.

#### A. Class I – Deterministic Discrete Swapping of Singular Vectors

For a 2x2 channel, $\mathbf{u}_1$ and $\mathbf{u}_2$ will swap place instantly and simultaneously with $\mathbf{v}_1$ and $\mathbf{v}_2$. In this instance, the magnitude of all the vector elements will be equal because this normally happens when a singular vector swap occurs. The singular values $s_i$ are constant. The one fundamental change that happens to the singular vectors when they swap is that they face phase inversions. Thus the vectors can be manipulated through a bipolar square wave generator:

$$\mathbf{U} = \begin{pmatrix} 1/\sqrt{2} & -\text{sgn}(\sin \omega t)/\sqrt{2} \\ \text{sgn}(\sin \omega t)/\sqrt{2} & 1/\sqrt{2} \end{pmatrix} \quad (A.1)$$

$$\mathbf{V} = \begin{pmatrix} 1/\sqrt{2} & \text{sgn}(\sin \omega t)/\sqrt{2} \\ -\text{sgn}(\sin \omega t)/\sqrt{2} & 1/\sqrt{2} \end{pmatrix} \quad (A.2)$$

where sgn() function returns the polarity of the sine wave input. The degree of "stress" can be altered by adjusting the frequency, $\omega$. This model can be applied to a 4x4 MIMO channel in a similar manner:

$$\mathbf{U} = \begin{pmatrix} z_a & z_c & z_c & z_a \\ z_b & z_b & z_b & z_b \\ z_b & z_b & z_d & z_d \\ z_a & z_c & z_a & z_c \end{pmatrix}; \quad \mathbf{V} = \mathbf{U}^T; \quad \begin{matrix} z_a = \tfrac{1}{2}\text{sgn}(\sin \omega t) \\ z_b = \tfrac{1}{2} \\ z_c = -\tfrac{1}{2}\text{sgn}(\sin \omega t) \\ z_d = -\tfrac{1}{2} \end{matrix} \quad (A.3)$$

### B. Class II – Deterministic Sinusoidal Change in Singular vectors

The change in singular vector elements **U** are gradual with a sinusoidal behavior, while **V** are constant. Unitary and orthogonal behavior is always retained and the frequency, $\omega$, at which the elements change can be adjusted. The singular values are also constant. Therefore:

$$\mathbf{U} = \begin{pmatrix} \sin \omega t & -\cos \omega t \\ \cos \omega t & \sin \omega t \end{pmatrix} \quad (A.4)$$

$$\mathbf{V} = \begin{pmatrix} 1/\sqrt{2} & -1/\sqrt{2} \\ 1/\sqrt{2} & 1/\sqrt{2} \end{pmatrix} \quad (A.5)$$

and another suitable solution can be applied to 4x4 MIMO:

$$\mathbf{Z} = \begin{pmatrix} z_a & -z_a & z_a & -z_a \\ z_b & -z_b & -z_b & z_b \\ z_a & z_a & -z_a & -z_a \\ z_b & z_b & z_b & z_b \end{pmatrix}; \quad \mathbf{U} = \mathbf{Z}\begin{cases} z_a = \tfrac{1}{2}\sin \omega t \\ z_b = \tfrac{1}{2}\cos \omega t \end{cases} \quad \mathbf{V} = \mathbf{Z}\begin{cases} z_a = \tfrac{1}{2} \\ z_b = \tfrac{1}{2} \end{cases} \quad (A.6)$$

### C. Class III – Semi-Deterministic Variable Singular vector Phases

This model aims to observe the effects of both continuously changing and randomly changing phase, while $|u_{ij}|$, $|v_{ij}|$ and $s_i$ are constant. For the 2x2 case, the matrix **V** is the same as that defined in equation (A.5), while equation (A.7) shows the phase in each element of **U** changes in a sinusoidal form but orthogonal singular vectors are maintained. If desired, it is also possible to apply a random phase, $\theta$ (thus becoming semi deterministic). Stress can be increased on the channel with the frequency, $\omega$.

$$\mathbf{U} = \begin{pmatrix} u_{\sin} & -u_{\sin} \\ u_{\cos} & u_{\cos} \end{pmatrix}; \quad (A.7)$$

$$u_{\sin} = \frac{e^{j\pi \sin(\omega t + \theta)}}{\sqrt{2}}; \quad u_{\cos} = \frac{e^{j\pi \cos(\omega t + \theta)}}{\sqrt{2}}$$

This model is also possible to implement for 4x4 MIMO using equation (A.6) for the **V** singular vectors. However, it is necessary to use a Householder transform as described in equations (A.12) to (A.14) to generate the **U** singular vectors. Furthermore a starting seed similar to (A.6) is required and a variable first column vector, $u_1$ as follows:

$$u_1 = \begin{pmatrix} u_{\sin} \\ u_{\cos} \\ u_{\sin} \\ u_{\cos} \end{pmatrix}; \quad \begin{aligned} u_{\sin} &= \frac{e^{j\pi \sin(\omega t + \theta)}}{2} \\ u_{\cos} &= \frac{e^{j\pi \cos(\omega t + \theta)}}{2} \end{aligned} \quad (A.8)$$

### D. Class IV – Semi-Deterministic Singular vectors Reflecting a Rayleigh Fading Environment

If the singular vectors are to behave in a way that reflects a 2x2 MIMO channel behavior in a Rayleigh fading environment, then it is necessary that the singular vectors are generated as random variables with memory defined by for example a Classical Doppler spectrum [29]. One possible method is to generate two random variables with memory, $a_1$ and $a_2$ using a ring scatterer [31] for $N_s$ scatterers at angle $\phi_i$ and random phase $\theta_i$. With sample factor, $S_f$

$$a_1(n) = \sum_{i=1}^{N_S} e^{j\left(\frac{2\pi n \sin \phi_i}{S_f} + \theta_{1i}\right)} \quad (A.9)$$

$$a_2(n) = \sum_{i=1}^{N_S} e^{j\left(\frac{2\pi n \sin \phi_i}{S_f} + \theta_{2i}\right)} \quad (A.10)$$

A suitably high enough value of $N_s \geq 6$ should be used [33][34]. A unitary vector, $u_1$ can now be created by applying the two variables as follows:

$$\mathbf{u}_1 = \begin{pmatrix} \dfrac{a_1}{\sqrt{|a_1|^2 + |a_2|^2}} \\ \dfrac{a_2}{\sqrt{|a_1|^2 + |a_2|^2}} \end{pmatrix} \quad (A.11)$$

The same process can be similarly achieved for singular vector $v_1$. For a 4x4 MIMO system, four independent variables $a_1$ to $a_4$ could be generated and produced to make a 1x4 singular vector in the same manner. The next challenge is to create the remaining singular vectors such that they are orthogonal and also unitary. This can be achieved for any size of singular vector matrix using the Householder transform. The simple application of a Householder transform $\mathbf{T}_u(t_n)$ (as defined in the case of **U**) from state (time step) $t_n$ to $t_{n-1}$ as [32]:

$$\mathbf{U}(t_n) = \mathbf{T}_u(t_n)\mathbf{U}(t_{n-1}) \quad (A.12)$$

where the transform matrix can be made based on the identity matrix **I** and the information of the previous state of the singular vector $\mathbf{u}_1$:

$$\mathbf{T}_u(t_n) = \mathbf{I} - \frac{\mathbf{w}\mathbf{w}^H}{\mathbf{w}^H \mathbf{u}_1(t_{n-1})} \quad (A.13)$$

$$\mathbf{w} = \mathbf{u}_1(t_{n-1}) - \mathbf{u}_1(t_n) \quad (A.14)$$

Applying this method will ensure that the transition made by the singular vectors is going from one legal state to the next. The same method is applied to $v_1$, which means that when the first singular vector has changed state, all the others can follow. As $\sum \lambda_i$ is constant, any mis-tracking of dominant modes or total mode gain is easily detectable. However, it is necessary that an initial state for the full orthogonal singular vector matrix $[\mathbf{U}(t_0), \mathbf{V}(t_0)]$ is set as a starting seed. This can be easily created as a quick SVD of a single sample.

APPENDIX B – FEATURES VS NUMERICAL COMPLEXITY ASPECTS OF CLASS V MODEL

The established approach to investigate and test eigen domain operation is to construct the physical channel links first followed by an SVD (e.g. LAPACK [28]) and hope for the desired eigen domain representation. If not achieved, a new iteration cycle is necessary to arrive at the desired output. Physical MIMO channels can be typically constructed using physical scatter layouts from which it is possible to assemble the multipath links 'piece by piece' or use a stochastic generator. Applying the LAPACK algorithm to such a MIMO channel will output the eigenmodes in sorted order of diversity. The class V model, on the other hand, has by design an inherent tracking of the natural path of each eigenmode. This is equivalent to mode untangling [23] of a LAPACK output as illustrated by a 2x2 channel case in Figure 7. Here the singular values of the class V model are shown to be consistent with the untangled case, while singular value swaps occur in the LAPACK case, notably at snapshot numbers 12 and 37. Singular vector rotation transients also occur at these points corresponding to a singular vector swap, though these changes are hidden in the physical channel dynamics. If selection diversity was performed on the singular values of the class V model, they would yield the same representation as the LAPACK algorithm outputs.

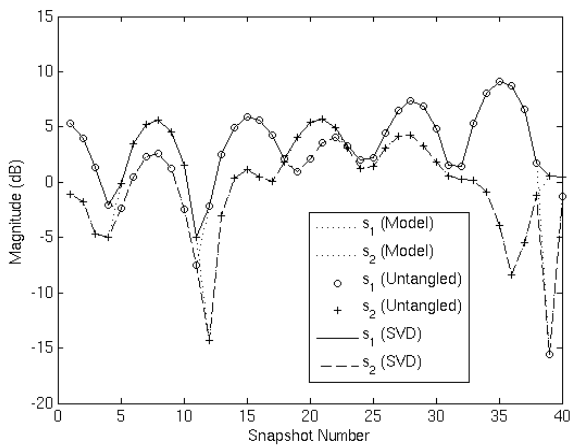

**Figure 7 – Comparison of time variant singular values in a class V 2x2 MIMO model with the untangled singular values with the assembled and re-decomposed singular values using LAPACK SVD algorithm**

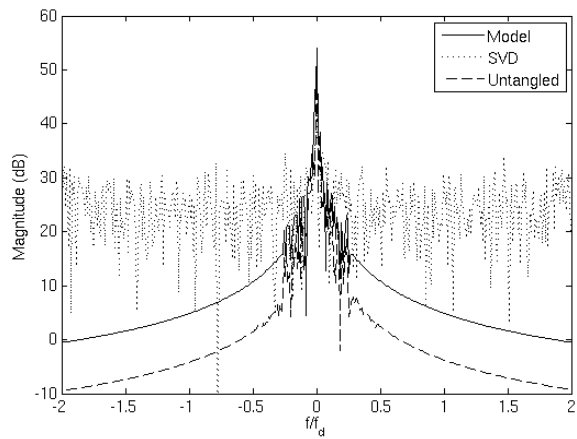

**Figure 8 - Comparison of the Doppler spectra of singular vector element $u_{11}$ between the class V model and the LAPACK singular value decomposition as well as the spectra following untangling**

The phase transients caused by singular value and corresponding singular vector swaps will cause the Doppler spectra of the singular vectors to be of an infinite range as illustrated in Figure 8 for the SVD or LAPACK case. Here the spectral extension for the LAPACK output is comparably wider than the class V/untangled case. This illustrates demands on Nyquist updating rates of eigenmode based MIMO communication devices, which are not constrained by a finite maximum Doppler shift. This is the reason updating singular vectors is more critical than updating physical channel states. A further feature of untangled singular values is that they hold the same cumulative distribution as illustrated by a class V 2x2 model in Figure 9. The SVD or LAPACK version has two separate distributions while for the Class V/untangled modes, the two distributions are overlapping and show the same gradient.

The class V model numerical complexity is of the order of a single iterative loop to run virtual Doppler noise generators for a 2x2 channel while two single loops are required for higher order channels (to use the Householder transform). This is favorable compared to physical modeling and the use of LAPACK and an untangling algorithm, which requires several nested iterative loops [23] reducing several seconds of runtime to a fraction of a second. To form the equivalent of LAPACK singular values, the additional selection diversity is a small extra effort on the class V mode.

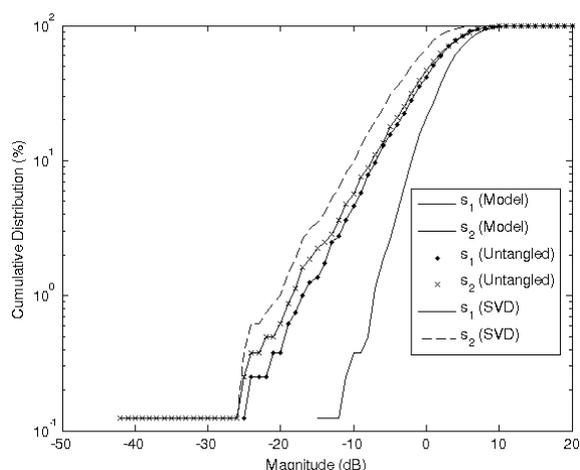

**Figure 9 - Comparison of the singular value cumulative distribution for both Class V/untangled and LAPACK SVD decompositions**

ACKNOWLEDGEMENTS

Many thanks go to Petar Popovski and Jesper Ødum Nielsen for their helpful comments in reviewing this paper. Acknowledgement is also given to David W. Browne for provision of code to generate untangled eigenmodes used to compare with results generated for this model.

VII. REFERENCES

[1] J. H. Winters, "On the capacity of radio communications systems with diversity in Rayleigh fading environments", IEEE Journal on Selected Areas in Communication, vol. 5, pp871-878, June 1987.
[2] J. B. Andersen, "Propagation Aspects of MIMO Channel Modeling", Chapter 1 of H. Bölcskei, D. Gesbert, C. B. Papadias, A.-J. van der Veen, "Space-time Wireless Systems: From Array Processing to MIMO Communications", *Cambridge*, 2008.
[3] J. W. Wallace, M. A. Jensen, "Modeling the Indoor MIMO Wireless Channel", IEEE Transactions on Antennas and Propagation, vol. 50, no. 5, May 2002.
[4] J. W. Wallace, M. A. Jensen, "Modeling the Indoor MIMO Wireless Channel", IEEE Transactions on Antennas and Propagation, vol. 50, no. 5, May 2002.
[5] D. Gesbert, H. Bölcskei, D. A. Gore, A. J. Paulraj, "Outdoor MIMO Wireless Channels: Models and Performance Prediction", IEEE Transactions on Communications, vol. 50, no. 12, December 2002.
[6] H. Xu, D. Chizhik, H. Huang, R. Valenzuela, "A Generalized Space-Time Multiple-Input Multiple-Output (MIMO) Channel Model", IEEE Transactions on Wireless Communications, vol. 3, no. 3, May 2004.
[7] T. Brown, E. DeCarvalho, P. Kyritsi, "Practical Guide to the MIMO Radio Channel", Wiley, 2012.
[8] M. A. Jensen, J. W. Wallace, "A review of antennas and propagation for MIMO wireless communications", IEEE Transactions on Antennas and Propagation, vol. 52, no, 11, November 2004, pp2810-2824.
[9] K. Yu, B. Ottersten, "Models for MIMO Propagation Channels: A Review", Wiley Wireless Communications and Mobile Computing, vol. 2, no. 7, November 2002, pp653-666.
[10] M. Shafi, M. Zhang, A. L. Moustakas, P. J. Smith, A. F. Molisch, F. Tufvesson, S. H. Simon, "Polarized MIMO channels in 3-D: models, measurements and mutual information", IEEE Journal on Selected Areas in Communications, vol. 24 no. 3, March 2006, pp514-527.
[11] D-S Shiu, G. J. Foschini, M. J. Gans, J. M. Kahn, "Fading Correlation and its Effect on the Capacity of Multielement Antenna Systems", IEEE Transactions on Communications, vol. 48, no. 3, March 2000.
[12] J. P. Kemoral, L. Schumacher, K. I. Pedersen, P. E. Mogensen, F. Frederiksen, "A Stochastic MIMO Radio Channel Model with Experimental Validation", IEEE Journal on Selected Areas in Communications, vol. 20, issue 6, pp1211-1226, August 2002.
[13] W. Weichselberger, M. Herdin, H. Özcelik, E. Bonek, "A Stochastic MIMO Channel Model with Joint Correlation of Both Link Ends", IEEE Transactions on Wireless Communications, vol. 5, no. 1, January 2006.
[14] A.M. Sayeed, "Deconstructing Multiantenna Fading Channels", IEEE Transactions on Signal Processing, vol. 50, no. 10, pp. 2563 – 2579, October 2002.
[15] N. Costa, S. Haykin, "A Novel Wideband MIMO Channel Model and Experimental Validation", IEEE Transactions on Antennas and Propagation, vol. 56, no. 2, February 2008, pp550-562.
[16] F. Schäfer, M. Stege, C. Michalke, G. Fettweis, "Efficient Tracking of Eigenspaces and its Application to Eigenbeamforming", IEEE International Symposium on Personal, Indoor and Mobile Radio Communications (PIMRC), vol. 3, 2003, pp2847-2851.
[17] T. W. C. Brown, P. C. F. Eggers, M. Katz, "Experimentation of Eigenvector Dynamics in a Multiple Input Multiple Output Channel", Wireless Personal Multimedia Communications Conference, 2005, September 2005, vol. 1, pp406-410.
[18] T. J. Willink, "Wide-Sense Stationarity of Mobile MIMO Radio Channels", IEEE Transactions on Vehicular Technology, vol. 57, no. 2, 2008, pp704-714.
[19] J. W. Wallace, M. A. Jensen, "Time-Varying MIMO Channels: Measurement, Analysis, and Modeling", IEEE Transactions on Antennas and Propagation, vol. 54, no. 11, November 2006, pp3265-3273.
[20] M. Herdin, N. Czink, H. Ozcelik, E. Bonek, "Correlation matrix distance, a meaningful measure for evaluation of non-stationary MIMO channels", IEEE Vehicular Technology Conference, vol. 1, 2005, pp136-140.
[21] M. Herdin, E. Bonek, "A MIMO Correlation Matrix based Metric for Characterizing Non-Stationarity", Proceedings IST Mobile and Wireless Communications Summit, Lyon, France, June 2004.
[22] I. Viering, H. Hofstetter, W. Utschick, "Validity of Spatial Covariance Matrices over Time and Frequency", IEEE Global Telecommunications Conference, vol. 1, 2001, pp851-855.
[23] D. W. Browne, M. W. Browne, M. P. Fitz, "Untangling the SVD's of Random Matrix Sample Paths", Allerton Conference on Communication, Control and Computing, 2006.
[24] R. G. Vaughan, J. Bach Andersen, "Channels, Propagation and Antennas for Mobile Communications", IEE Press, 2002.
[25] L. Wood and W. S. Hodgkiss, "A Reduced-Rank Eigenbasis MIMO Channel Model", Wireless Telecommunications Symposium, pp78-83 2008.
[26] A. Alcocer–Ochoa, V. Kontorovitch, R. Parra–Michel , "A Wideband MIMO Channel Model based on Structured Vector Modes using Orthogonalization", International Conference on Wireless Communications, Networking and Mobile Computing, pp1-5, 2008.
[27] Y. Zhang, O. Edfors, P. Hammarberg, T. Hult, X. Chen, S. Zhou, L. Xiao, J. Wang , "A General Coupling-Based Model Framework for Wideband MIMO Channels", IEEE Transactions on Antennas and Propagation, vol. 60, no. 2, pp574-586, February 2012.
[28] E. Anderson, Z. Bai, C. Bischof. S. Blackford, J. Demmel, J. Dongarra, J. Du Croz, A. Greenbaum, S. Hammarling, A. McKenney, D. Sorensen, "LAPACK Users Guide", Third Edidtion, Siam, 1999.
[29] S. Saunders, "Antennas and Propagation for Wireless Communication Systems", Wiley, UK, 1999.
[30] J. Bach Andersen, "Array Gain and Capacity for Known Random Channels with Multiple Element Arrays at Both Ends", IEEE Journal on Selected Areas in Communications, vol. 18, no. 11, November 2000.
[31] M. Patzold, "Mobile Radio Channels", 2nd Edition, Wiley, 2011.
[32] K-L. Chung, W-M. Yan, "The Complex Householder Transform", IEEE Transactions on Signal Processing, vol. 45, no. 9, pp2374-2376, September 1997.
[33] W. C. Jakes, "Microwave Mobile Communications", IEEE Press, 1974.
[34] R. H. Clarke, "A Statistical Theory of Mobile-Radio Reception", Bell System Technical Journal, vol. 47, issue 6, pp957–1000, July-August 1968.